\documentclass[9pt]{article}
\usepackage{spconf,amsmath,graphicx,hyperref}
\usepackage{amsfonts,amsthm}
\usepackage{amssymb}
\usepackage{pifont}
\usepackage{algorithm}
\usepackage{algpseudocode}
\usepackage{threeparttable}
\usepackage{makecell}
\usepackage{colortbl}
\usepackage{multirow}
\usepackage{enumerate}
\usepackage{subfig}
\usepackage{comment}
\usepackage{enumitem}
\newtheorem{theorem}{\bf Theorem}
\newtheorem{definition}{\bf Definition}

\usepackage{acronym}
\usepackage{xspace}
\usepackage{xcolor}

\newcommand{\ie}{i.e.,\xspace}
\newcommand{\eg}{e.g.,\xspace}

\newcommand{\eat}[1]{}

\acrodef{iot}[IoT]{Internet of Things}

\newcommand{\mapping}[1]{}
\newcommand{\code}{\url{https://github.com/bkjod/FedSVA_Shapley}}

\newcommand{\tool}{{FedSVA}\xspace}
\acrodef{sgd}[FedSGD]{Federated Stochastic Gradient Descent}

\acrodef{avg}[FedAvg]{Federated Averaging}

\newcommand{\heading}[1]{\vspace{4pt}\noindent{\underline{\textsc{#1}}}}

\title{Towards Explainable Privacy Preservation in Federated Learning via Shapley Value-Guided Noise Injection}
\name{Yunbo Li, Jiaping Gui*\thanks{*Corresponding authors: \{jgui, wuyue\}@sjtu.edu.cn}, Yue Wu*}
\address{School of Computer Science, Shanghai Jiao Tong University\\
	}
\begin{document}
%
\maketitle

\begin{abstract}

This paper proposes \tool, an explainable differential privacy (DP) mechanism for federated learning (FL) that dynamically calibrates noise injection based on the privacy contribution of attributes via Shapley Values. Unlike heuristic DP methods, \tool quantifies each attribute’s influence on model training and adjusts noise accordingly, providing rigorous privacy guarantees while minimizing utility loss. Theoretical analysis confirms convergence and DP properties. Experiments on CIFAR-10 and FEMNIST show state-of-the-art privacy-utility trade-offs and robust defense against reconstruction attacks. Code is available at: \code.


\end{abstract}
\section{Introduction}
\label{sec:intro}

Federated Learning (FL)~\cite{mcmahan2017communication} has emerged as a promising paradigm that enables multiple participants to collaboratively train machine learning models without sharing their local data. By keeping data on-device and only exchanging model updates, FL inherently reduces the risk of raw data exposure. Despite this advantage, recent studies have revealed that FL systems remain susceptible to sophisticated privacy attacks~\cite{hitaj2017deep}, such as model inversion~\cite{zhu2019deep} and membership inference~\cite{shokri2017membership}, wherein adversaries can reconstruct sensitive training data or infer private attributes from the exposed model parameters.

To counter these threats, Differential Privacy (DP)~\cite{dwork2014algorithmic} has been widely adopted as a rigorous mathematical framework for privacy preservation. By carefully injecting calibrated noise during the training process, DP prevents an adversary from determining whether any particular sample was included in the training set~\cite{wei2020federated,fu2022adap}. Existing DP mechanisms in FL often determine the noise scale through heuristic or indirect factors, such as training progress~\cite{wei2023securing,fukami2024dp} or time-varying schedules~\cite{yuan2023privacy,yuan2023amplitude}. However, these approaches lack a clear and interpretable connection to the actual privacy contribution of the underlying data, making it difficult to explain the rationale behind specific noise levels. This opacity can lead to either excessive noise injection that degrades model utility, or insufficient noise that leads to inadequate protection~\cite{zhou2022pflf}.

To address this gap, we propose \tool, a novel and explainable DP-based privacy protection mechanism for FL. Our approach is grounded in the key insight that the amount of injected noise should be explicitly guided by the contribution of private attributes to the model’s performance. To quantify this contribution, we leverage Shapley Values~\cite{lundberg2017unified}, a concept from cooperative game theory, to evaluate the influence of each attribute on the local learning process. Based on this valuation, \tool dynamically adjusts the noise injection in each training round, ensuring that privacy protection is both effective and efficient.

The main contributions of this work are as follows:

\begin{itemize}[leftmargin=*, noitemsep, topsep=0pt]
	\item We introduce \tool, the first FL privacy mechanism that integrates Shapley Values with differential privacy to provide an explainable noise injection strategy.
	\item We provide a rigorous theoretical analysis of \tool, including formal privacy guarantees and a convergence analysis for the federated optimization process.
	\item Through extensive experiments on standard benchmarks and models, we demonstrate that \tool achieves a superior privacy-utility trade-off compared to state-of-the-art methods, and effectively defends against real-world data reconstruction attacks.
\end{itemize}

\section{Background and Problem Formulation}
\label{sec:background}

\subsection{Federated Learning and Privacy Vulnerabilities}
Federated learning involves a central server $S$ and $M$ clients $\{\mathcal{C}_i\}_{i=1}^M$, each with local dataset $\mathcal{D}_i$ of size $n_i$. The goal is to minimize $F(w) = \sum_{i=1}^m \frac{n_i}{n} F_i(\mathcal{D}_i;w_g)$ where $n = \sum_{i=1}^M n_i$, $F_i$ is the local loss, and $w_g$ represents global parameters. Each global round $t$, a subset of $K$ clients (rate $r=K/M$) performs $E$ local epochs to produce $w_i^t$. The server then aggregates parameters via FedAvg (or FedSGD): $w_g^t = \sum_{i=1}^K \frac{n_i}{n} w_i^t$, and distributes $w_g^t$ to participants.

While FL mitigates direct data exposure, recent studies have revealed that shared model parameters can inadvertently leak sensitive information about the training data~\cite{hitaj2017deep, zhu2019deep}. Adversaries, even those behaving as Honest-but-Curious (HBC) participants who faithfully train their local models, can launch privacy attacks such as membership inference~\cite{shokri2017membership} or training data reconstruction~\cite{hitaj2017deep} by analyzing the global model updates. These vulnerabilities necessitate robust privacy-preserving mechanisms within FL frameworks.

\subsection{Differential Privacy in Federated Learning}
Differential Privacy (DP) has emerged as a rigorous, mathematical standard for privacy protection~\cite{wei2020federated}. In FL, DP techniques typically involve injecting calibrated noise into the local model updates before they are shared with the server, thereby obfuscating the contribution of any single data point and providing a quantifiable privacy guarantee $(\epsilon, \delta)$.

However, a key challenge lies in determining \textit{how much} noise to inject. Existing DP-FL schemes often determine the noise scale based on factors indirectly related to privacy, such as training progress~\cite{yuan2023amplitude} or the number of participating clients~\cite{xie2022securing}. This indirect approach makes it difficult to explain the relationship between the injected noise and the actual privacy risk posed by the specific data being protected. Consequently, these methods can suffer from a suboptimal privacy-utility trade-off, leading to either excessive noise degradation of model utility or insufficient noise, increasing the risk of privacy breaches.

\subsection{Problem Formulation: Explainable Noise Scheduling}
The core problem we address is the \textbf{lack of explainability and precision in dynamic noise scheduling for DP in FL}. The goal is to move beyond heuristic or indirect noise calibration towards a principled method where the amount of injected noise is explicitly guided by the \textit{quantifiable contribution of private attributes} within a client's local dataset.

We consider a standard FL system with a central server and $M$ clients. A subset of clients are HBC adversaries aiming to reconstruct other clients' private data. Each benign client $i$ possesses a local dataset $\mathcal{D}_i$. We allow clients to orthogonally partition their data into subsets based on attributes (\eg $\mathcal{D}_i^p$ for private attributes and $\mathcal{D}_i^a$ for non-private attributes). The objective is to design a client-side DP mechanism $\mathcal{M}$ that, for each training round $t$:

\begin{enumerate}[leftmargin=*, noitemsep, topsep=0pt]
	\item \textbf{Estimates} the contribution rate $R_i^p$ of the private data subset $\mathcal{D}_i^p$ to the local training process.
	\item \textbf{Dynamically adjusts} the noise variance $\sigma_i^2$ proportional to $R_i^p$, ensuring a stricter ($(\epsilon, \delta')$-DP) guarantee.
	\item \textbf{Preserves utility} by minimizing unnecessary noise injection when the privacy risk is low, thereby achieving a superior and explainable privacy-utility trade-off compared to state-of-the-art methods.
\end{enumerate}

\begin{figure}[!t]
    \centering
    \includegraphics[width=1\linewidth]{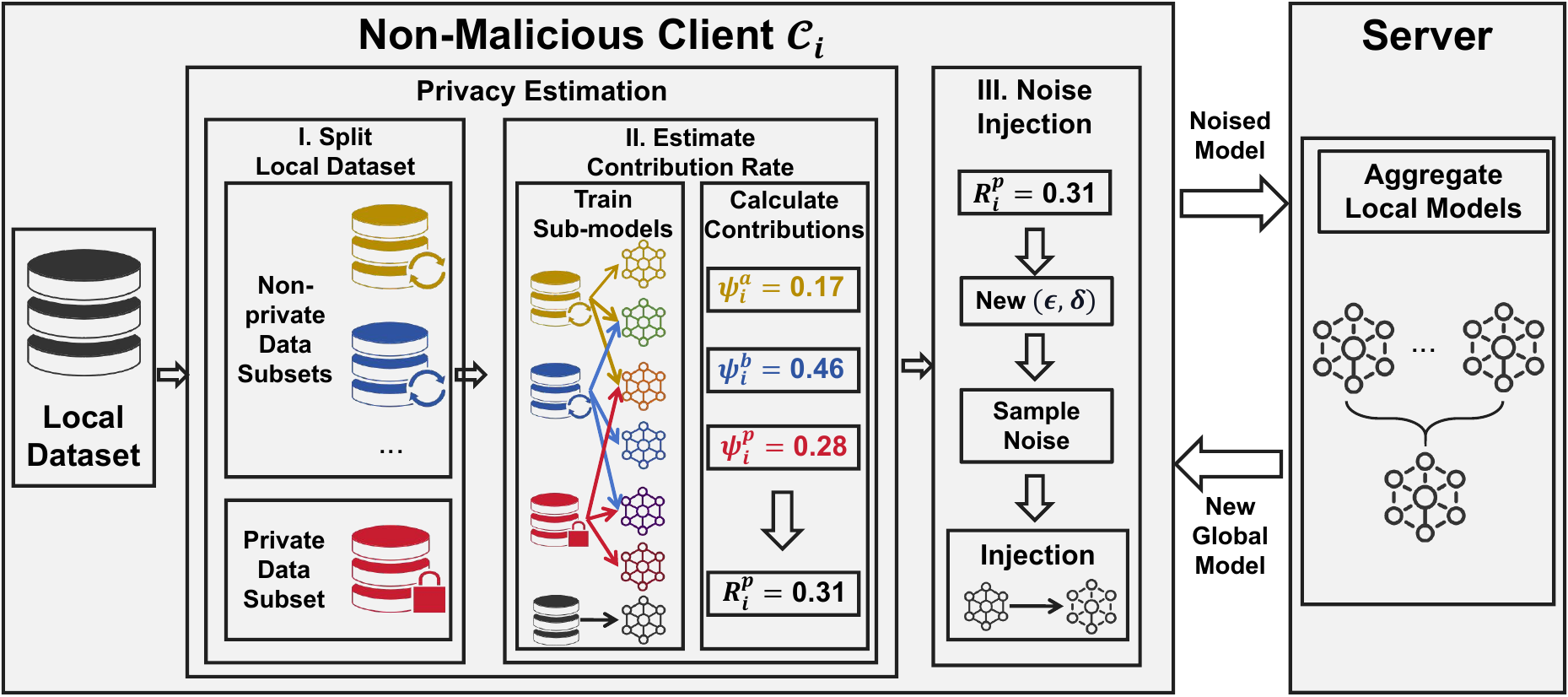}
    \caption{System overview of \tool for a single client. The client computes the privacy contribution rate $R_i^p$ using the Privacy Estimation (PE) module, determines the noise scale $\sigma_i$ based on $R_i^p$ in the Noise Injection (NI) module, and uploads the perturbed update $\widetilde{w}_i^t = w_i^t + n_i$ to the server.}
    \label{fig:system_overview}
\end{figure}

\begin{algorithm}[t]
\footnotesize
\caption{The Workflow of \tool}\label{algo_workflow}
\begin{algorithmic}[1]
\State Initialize $w_g$, clients $\mathcal{C}$, activation rate $r$, local epochs $E$, rounds $T$
\For{$t=1,\dots,T$}
\State Server selects $K=r|\mathcal{C}|$ clients, broadcasts $w_g^t$.
\State Each selected client trains $E$ epochs $\rightarrow w_i^t$.
\State \textbf{PE:} estimate $R_i^p(t)$ via Shapley-based attribute contributions.
\State \textbf{NI:} calculate $\sigma_i$ by $R_i^p(t)$ and sample $n_i\sim\mathcal{N}(0,\sigma_i^2 \mathbb{I}^d)$.
\State Client uploads $\widetilde{w}_i^t=w_i^t+n_i$.
\State Server aggregates by FedAvg.
\EndFor
\end{algorithmic}
\end{algorithm}

\section{Design of \tool}
\label{sec:3}

To realize \tool, each participating client is equipped with two key modules: (i) Privacy Estimation (PE), which uses Shapley Values to quantify the contribution rate of designated private attributes, and (ii) Noise Injection (NI), which maps this contribution rate to the Gaussian noise scale for perturbing local updates, yielding formal DP guarantees. Figure~\ref{fig:system_overview} illustrates the overall client-side workflow of \tool, and Algorithm~\ref{algo_workflow} details the complete federated training process.


\subsection{Privacy Estimation (PE) Module} 
The PE module quantifies the contribution of private data attributes to the local model's utility using Shapley Values~\cite{lundberg2017unified}. Each client first partitions its local dataset $\mathcal{D}_i$ into subsets corresponding to $N$ attributes $\mathcal{A}=\{a_o\}_{o=1}^N$ as $\{D^a_i\}_{a\in \mathcal{A}/\{p\}}\cup \{D^p_i\}$, where $p$ denotes privacy attributes. 

During global round $t$, non-malicious clients enumerate the attribute power set $\Theta(\mathcal{A})$. For each $\theta \in \Theta(\mathcal{A})$, \tool constructs a new training dataset $\mathcal{D}^{\theta}_i = \bigcup_{a\in\theta}\mathcal{D}_i^a$ and fine-tunes auxiliary model $w_i^{t,\theta}$ from $w_g^t$. Subsequently, \tool evaluates the utility $U(w_i^{t,\theta})$ via local validation accuracy. The Shapley Value for an attribute $a$ is computed as $\psi^a_i = \frac{1}{|\mathcal{A}|!}\sum_{\pi \in \Pi(\mathcal{A})}[U(w_i^{t,g(P_a^{\pi}\cup\{a\})}) - U(w_i^{t,g(P_a^{\pi})})]$, where $g: \{\mathcal{A}_s|\mathcal{A}_s \subseteq \mathcal{A} \land \mathcal{A}_s \neq \phi\} \to \Theta(\mathcal{A})$ maps all non-empty subsets of $\mathcal{A}$ to unique combinations in $\Theta(\mathcal{A})$. A larger $\psi^a_i$ indicates a higher contribution of the attribute $a$. After obtaining the Shapley Values for each attribute, \tool calculates $R_i^p = \frac{\psi_i^p}{\sum_{a\in\mathcal{A}/\{p\}}\psi_i^a+\psi_i^p}$ to quantify the contribution rate of privacy attributes. Note that for the extreme case where client $\mathcal{C}_i$ has $\mathcal{D}^p_i=\mathcal{D}_i$, the contribution rate satisfies $R_i^p = 1$ since we have $\psi_i^a = 0$ for $\forall a\in\mathcal{A}/\{p\}$. This implies that \tool remains effective even if the client cannot partition its local dataset according to the attribute or treats the entire dataset as private information.

\subsection{Noise Injection (NI) Module} 
The NI module dynamically adjusts the Gaussian noise scale based on $R_i^p$. It translates the contribution rate into a risk tolerance for privacy leakage, defined as $T_i^p = R_i^p - \ln \delta^2$, with a minimum enforced baseline $\beta_i > 0$ to ensure fundamental protection. According to the Gaussian mechanism of Differential Privacy~\cite{dwork2014algorithmic}, the noise standard deviation for the client's model update (with sensitivity $\Delta f_i = 2C/|\mathcal{D}_i|$ bounded by gradient clipping~\cite{wei2020federated}) is calibrated as $\sigma_i = \frac{2C\sqrt{\max(T^p_i, \beta_i)}}{\epsilon|\mathcal{D}_i|}$. The client subsequently samples a noise matrix with the same shape $d$ as the parameter matrix (\ie $n_i \sim \mathcal{N}(0, \sigma_i^2 \mathbb{I}^d)$) and uploads the perturbed update $\widetilde{w}_i^t = w_i^t + n_i$ to the server.

\subsection{Privacy and Convergence Analysis}
The privacy guarantee of \tool is formalized as follows. Given the privacy budget $(\epsilon, \delta)$, \tool provides an $(\epsilon, \delta')$-DP guarantee in a single round and an $(T\epsilon, T\delta')$-DP guarantee over $T$ rounds (based on the basic composition theorem), where $\delta' \leq 1.25e^{(-R_i^p/2)}\delta$. For clients with a low privacy contribution rate, specifically $R_i^p < \beta_i + \ln\delta^2$, the mechanism provides at least an $(\epsilon, \delta_0)$-DP guarantee with $\delta_0 \leq 1.25e^{(-\beta_i/2)}$. This result is derived from the privacy analysis of Gaussian DP~\cite{dwork2014algorithmic} by equating the risk tolerance $T_i^p$ to the term $2\ln(1.25/\delta')$ that governs the Gaussian noise scale.

The convergence of \tool is analyzed under standard assumptions~\cite{li2019convergence, lifedqs}: the global loss function $F(x)$ is $L$-smooth and $\rho$-weakly convex; the expected squared norm of local stochastic gradients is bounded by $G_c^2$; and the expected squared norm of the injected noise is bounded by $N_c^2$. Then, given a learning rate $\eta < 1/\rho$, the expected loss of the noised global model $\widetilde{w}_g^t$ after $t$ rounds is bounded by
$\mathbb{E}[F(\widetilde{w}_g^t)] - F^* \leq L\mathcal{U}^t\mathbb{E}[|w_g^0 - w^*|^2] + \frac{14}{\rho^2}(1-\mathcal{U})\mathcal{W} + 4\eta^2\mathcal{W} + LK^2N_c^2$,
where $F^*$ is the global minimum, $\mathcal{U} = 1 - \eta\rho$, and $\mathcal{W} = L(E-1)^2G_c^2$. The bound consists of a term decaying exponentially with $t$, terms related to the gradient norms, and a final term $LK^2N_c^2$ that quantifies the error introduced by aggregating client-level noise. This demonstrates that the convergence of \tool is constrained by the noise injected for privacy protection, with a quadratic dependence on the number of clients selected per round $K$.

\noindent \textbf{Proof Sketch.} (1) Global drift bound: By AM–GM inequality, Cauchy–Schwarz inequality, and Lemma 3 in~\cite{li2019convergence}, the global model drift satisfies $\mathbb{E}[||w_g^t  - w^*||^2]  \leq \mathcal{U}^{t}\mathbb{E}[||w_g^0 - w^*||^2]  + (\frac{14}{\rho}+4\eta)\eta (E-1)^2G_c^2$. (2) Noise aggregation bound: Expanding $||\sum_{i=1}^K p_in_i^t||^2$ and bounding cross terms gives $\mathbb{E}[||\sum_{i=1}^K p_i n_i^t||^2] \leq K^2N_c^2$. (3) Overall convergence bound: From $L$-smoothness and $\nabla F^*=0$, we have $\mathbb{E}[F(\widetilde{w}_g^t)] - F^* \leq \mathbb{E}[L||w_g^t  - w^*||^2] + \mathbb{E}[L||\sum_{i=1}^K p_i n_i^t||^2]$. Then, applying (1) and (2) yields the final conclusion.

\section{Evaluation}
\label{sec_eva}

\begin{table}[!t]
    \centering
    \caption{Accuracy comparison among different methods in two experimental scenarios.}
    \scriptsize
    \setlength{\tabcolsep}{3pt}
    \begin{threeparttable}
    \begin{tabular}{c|c|c c| c c}
    \hline
     \multirow{4}{*}{\makecell{Privacy \\ Budget}} & \multirow{4}{*}{Method} & \multicolumn{4}{c}{Experiment Scenario} \\ \cline{3-6}
    & & \multicolumn{2}{c|}{ResNet-18 @ CIFAR-10} & \multicolumn{2}{c}{CNN @ FEMNIST} \\\cline{3-6}
    & &\makecell{Best Acc.}  & \makecell{Final-5 \\Average Acc} & \makecell{Best Acc.}  & \makecell{Final-5 \\Average Acc.}\\ \cline{1-6}
    - & \textbf{FedAvg} &  \textbf{85.54 $\pm$ 1.22}  & \textbf{84.33 $\pm$ 0.51} & \textbf{79.22 $\pm$ 0.48} & \textbf{77.79 $\pm$ 0.13} \\\cline{1-6}

    \multirow{5}{*}{0.2} & NbAFL &  56.48 $\pm$ 5.38& 48.67 $\pm$ 2.39&  43.58 $\pm$ 2.12& 22.66 $\pm$ 3.32 \\
    & TV DP &  64.10 $\pm$ 3.76& 61.09 $\pm$ 2.18 &  53.88 $\pm$ 3.43 & 50.92 $\pm$ 1.06 \\
    & Sens DP &  65.09 $\pm$ 4.51& 63.14 $\pm$ 3.01 &  67.33 $\pm$ 5.87& 68.92 $\pm$ 0.09 \\
    & DPA DP &  67.83 $\pm$ 2.85 & 65.21 $\pm$ 1.43 &  70.79 $\pm$ 4.12& 69.57 $\pm$ 0.68 \\
    & \cellcolor[gray]{0.9}\textbf{\tool} & \cellcolor[gray]{0.9}\textbf{75.86 $\pm$ 0.37}& \cellcolor[gray]{0.9}\textbf{75.98 $\pm$ 2.71} & \cellcolor[gray]{0.9}\textbf{73.19 $\pm$ 0.56} &\cellcolor[gray]{0.9}\textbf{71.93 $\pm$ 0.12} \\\cline{1-6}

    \multirow{5}{*}{0.3} & NbAFL &  64.53 $\pm$ 2.18 & 62.70 $\pm$ 0.52 &  66.16 $\pm$ 1.37 & 62.49 $\pm$ 0.24 \\
    & TV DP &  73.42 $\pm$ 1.51 & 71.63 $\pm$ 0.45 &  71.35 $\pm$ 0.86& 69.22 $\pm$ 0.12 \\
    & Sens DP &  73.88 $\pm$ 4.22 & 71.65 $\pm$ 1.69&  71.22 $\pm$ 3.59& 71.03 $\pm$ 0.04\\
    & DPA DP &  75.05 $\pm$ 1.38 & 73.47 $\pm$ 1.01&  72.25 $\pm$ 1.77& 71.75 $\pm$ 0.14\\
    & \cellcolor[gray]{0.9}\textbf{\tool }& \cellcolor[gray]{0.9}\textbf{79.17 $\pm$ 0.68} & \cellcolor[gray]{0.9}\textbf{77.93 $\pm$ 0.75} & \cellcolor[gray]{0.9}\textbf{73.79 $\pm$ 0.52} & \cellcolor[gray]{0.9}\textbf{73.18 $\pm$ 0.47} \\\cline{1-6}

    \multirow{5}{*}{0.4} & NbAFL &  72.77 $\pm$ 0.84 & 70.35 $\pm$ 0.47&  71.33 $\pm$ 0.26 & 70.85 $\pm$ 0.10\\
    & TV DP &  76.64 $\pm$ 0.44 & 75.71 $\pm$ 0.28&  71.95 $\pm$ 0.79& 71.83 $\pm$ 0.05\\
    & Sens DP &  75.93 $\pm$ 0.98 & 74.87 $\pm$ 2.66 & 71.72 $\pm$ 0.68& 71.69 $\pm$ 0.22 \\
    & DPA DP &  78.09 $\pm$ 0.53 & 77.63 $\pm$ 0.80&  72.25 $\pm$ 0.39 & 71.92 $\pm$ 0.09 \\
    & \cellcolor[gray]{0.9}\textbf{\tool }& \cellcolor[gray]{0.9}\textbf{81.81 $\pm$ 0.58} & \cellcolor[gray]{0.9}\textbf{80.96 $\pm$ 0.61} & \cellcolor[gray]{0.9}\textbf{75.30 $\pm$ 0.29} & \cellcolor[gray]{0.9}\textbf{73.88 $\pm$ 0.48} \\\cline{1-6}

    \multirow{5}{*}{0.5} & NbAFL &  74.96 $\pm$ 2.33 & 73.02 $\pm$ 1.85 & 71.55 $\pm$ 0.45 & 71.28 $\pm$ 0.26\\
    & TV DP & 79.14 $\pm$ 0.14 & 78.15 $\pm$ 0.24& 72.63 $\pm$ 0.07 & 72.39 $\pm$ 0.06\\
    & Sens DP & 78.59 $\pm$ 1.37 & 77.24 $\pm$ 1.60& 72.21 $\pm$ 0.63 & 72.05 $\pm$ 0.14\\
    & DPA DP &  79.82 $\pm$ 0.67 & 79.24 $\pm$ 0.44&  73.18 $\pm$ 0.49 & 72.50 $\pm$ 0.19\\
    & \cellcolor[gray]{0.9}\textbf{\tool }& \cellcolor[gray]{0.9}\textbf{82.89 $\pm$ 0.11} & \cellcolor[gray]{0.9}\textbf{82.14 $\pm$ 1.93} & \cellcolor[gray]{0.9}\textbf{76.06 $\pm$ 0.16} & \cellcolor[gray]{0.9}\textbf{74.43 $\pm$ 0.12} \\\cline{1-6}
    \end{tabular}
    \begin{tablenotes}
        \footnotesize
        \item[*] ``Best Acc.'' represents the average highest test accuracy (\%) during training among all runs, and ``Final-5 Average Acc.'' represents the average accuracy (\%) during the final five training rounds among all runs. The values following $\pm$ indicate the standard variance.
      \end{tablenotes}
    \end{threeparttable}
    \label{tab:sota_utility}
\end{table}

\subsection{Experimental Setup}
\label{sec:exp-setup}
To evaluate \tool, we configure an FL system with 100 clients, including 10 Honest-but-Curious (HBC) clients. The maximum number of global rounds is $T=400$. We use a learning rate $\eta=0.1$, local epochs $E=2$, activation rate $r=10\%$, and gradient clipping norm $C=20$. The failure probability $\delta$ is set to the reciprocal of the local batch size. We conduct experiments on two benchmarks: (1) ResNet-18 on CIFAR-10 and (2) CNN on FEMNIST. HBC clients use an Auxiliary Classifier GAN (ACGAN) for reconstruction attacks. By default, \tool uses binary attribute partitioning ($N=2$). All experiments are averaged over five independent runs on an Intel(R) Xeon(R) E5-2620 v4 CPU and an NVIDIA GeForce GTX 1080 Ti GPU.

\subsection{Performance Evaluation}

\heading{Privacy-Utility Trade-off.} We compare \tool against four state-of-the-art (SOTA) DP-FL baselines: NbAFL~\cite{wei2020federated}, Time-Varying DP (TV DP)~\cite{yuan2023amplitude}, Sensitivity Re-estimation DP (Sens DP)~\cite{xue2023differentially}, and Dynamic Privacy Allocation LDP (DPA DP)~\cite{zhang2024dynamic}. Table~\ref{tab:sota_utility} reports the test accuracy under varying privacy budgets $\epsilon$. \tool consistently outperforms all baselines across all $\epsilon$ values on both datasets. For example, on CIFAR-10/ResNet-18 with $\epsilon=0.2$, \tool achieves a final-5 accuracy of $75.98\%$, significantly higher than DPA DP ($65.21\%$), Sens DP ($63.14\%$), TV DP ($61.09\%$), and NbAFL ($48.67\%$). As $\epsilon$ increases to 0.5, the accuracy of all methods improves, with \tool reaching $82.14\%$. This superior privacy-utility trade-off stems from \tool's precise, attribute-aware noise calibration, which minimizes unnecessary noise injection when the privacy risk is low.

\begin{figure}[!t]
\centering
\subfloat{\includegraphics[width=0.8\linewidth]{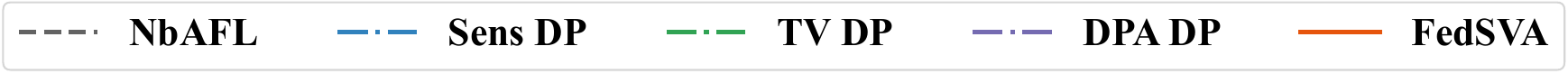}}
\vspace{-2ex}
\setcounter{subfigure}{0}
\subfloat[$\epsilon = 0.2$]{\includegraphics[width=0.35\linewidth]{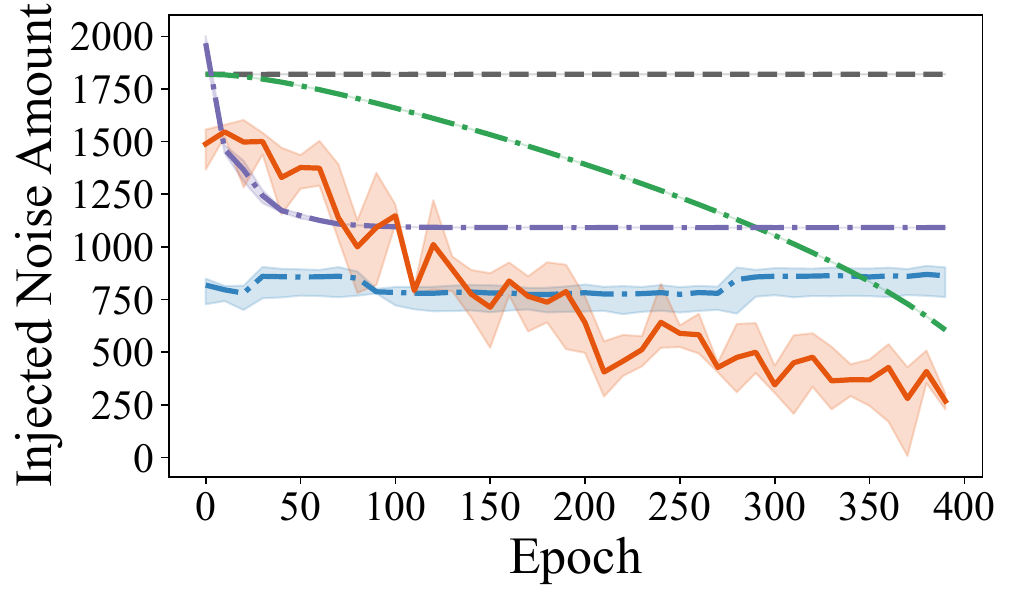}\label{fig_scale02_res_cifar}}
\hspace{1mm}
\subfloat[$\epsilon = 0.5$]{\includegraphics[width=0.35\linewidth]{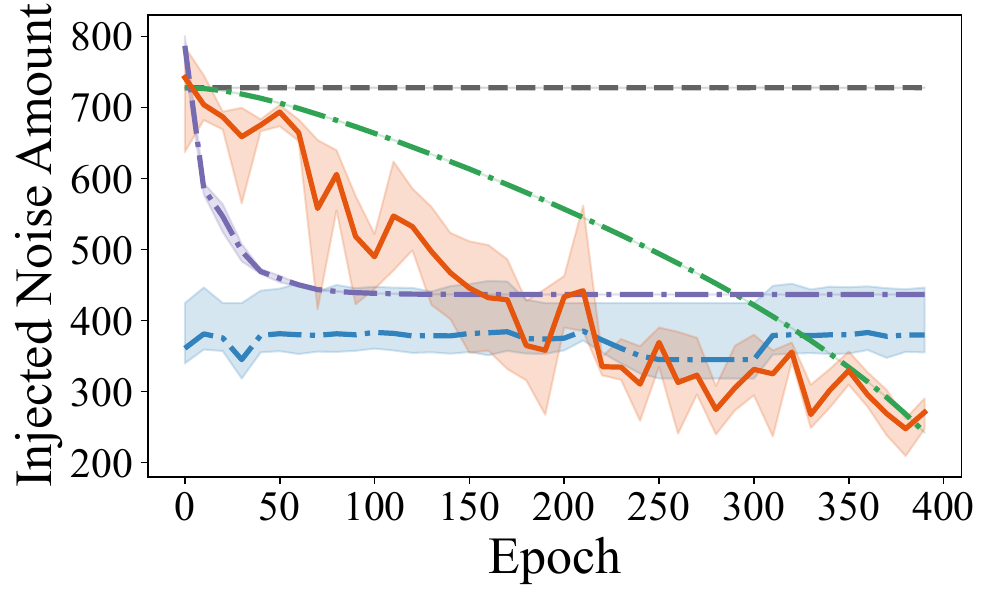}\label{fig_scale05_res_cifar}}
\hspace{1mm}
\vspace{-1ex}
\hspace{-1ex}
\subfloat[Accumulative Privacy Budget]{\includegraphics[width=0.7\linewidth]{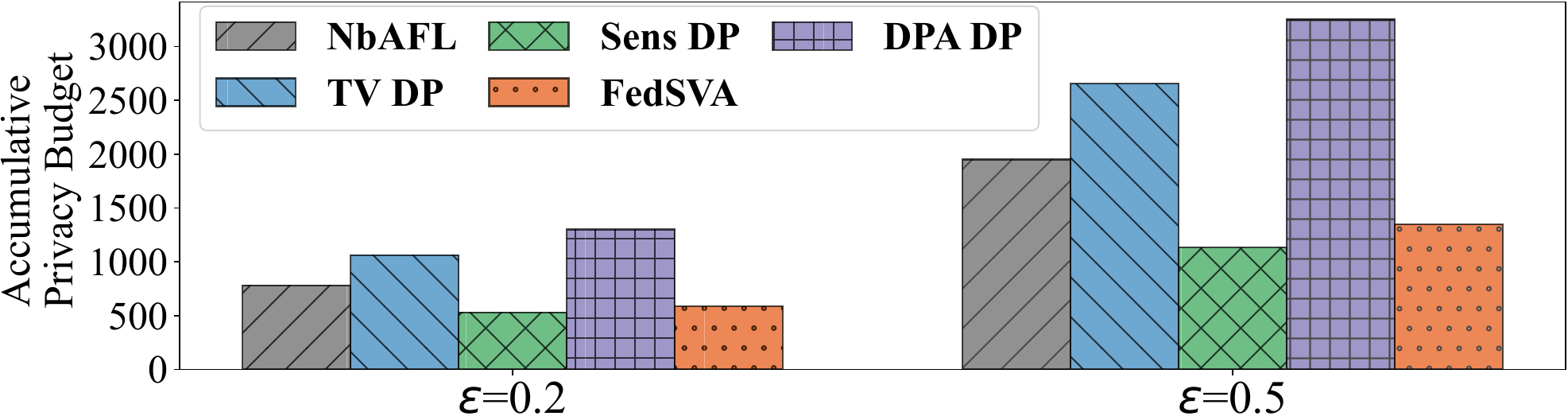}\label{fig_budget_res_cifar}}
\caption{Comparison of the injected noise (a-b) and the cumulative privacy budget (c) between \tool and baseline methods during the training of ResNet-18 on CIFAR-10.}
\label{fig_noise_res_cifar}
\end{figure}

\begin{figure}[!t]
	\centering
	\includegraphics[width=0.95\linewidth]{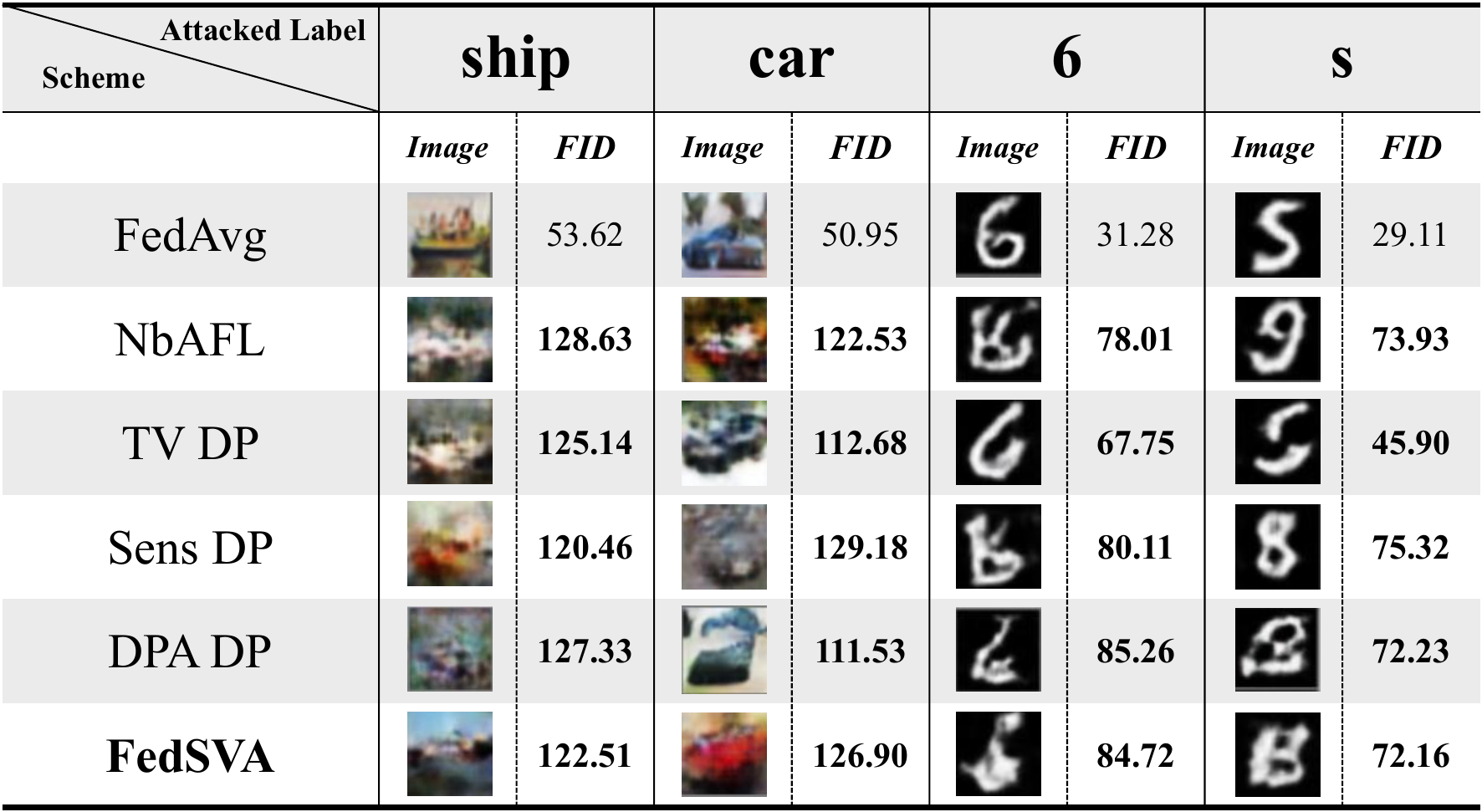}
	\caption{Representative reconstruction results.}
	\label{fig:att}
\end{figure}

\heading{Noise Injection and Privacy Budget Analysis.} Figures~\ref{fig_scale02_res_cifar}–\ref{fig_scale05_res_cifar} illustrate the cumulative $\mathcal{L}_2$ norm of noise injected per round by each method on CIFAR-10 for $\epsilon=0.2$ and $\epsilon=0.5$, respectively. \tool injects substantially less noise than the static NbAFL method. More importantly, compared to other dynamic baselines (TV DP, Sens DP, DPA DP), \tool's Shapley-value-driven strategy achieves a more precise noise schedule, injecting noise only when the private data contribution is significant. Figure~\ref{fig_budget_res_cifar} shows the cumulative privacy budgets during the overall training rounds, calculated using the sequential accountant based on the naive composition theorem~\cite{dwork2014algorithmic}. \tool consumes a smaller cumulative budget compared to several baselines, confirming its ability to provide strong privacy guarantees with a tighter overall bound, thanks to its explainable noise adjustment mechanism.

\heading{Defense Against Reconstruction Attacks.} We evaluate the defense capability against ACGAN-based reconstruction attacks~\cite{hitaj2017deep} under $\epsilon=0.3$. Reconstruction quality is measured by the Fréchet Inception Distance (FID) between 10,000 real and reconstructed images; a higher FID indicates better privacy protection. Figure~\ref{fig:att} shows representative reconstruction results and the average FID scores per class. The vanilla FedAvg model (no DP) is severely compromised, yielding easily recognizable reconstructions and low FID scores. \tool achieves baseline-comparable defense (149.07\% FID increase over FedAvg) despite reduced noise injection. This demonstrates that \tool's targeted noise injection effectively obfuscates private information without relying on excessive noise.

\heading{Impact of Attribute Partition Granularity ($N$).} 
We analyze the effect of partition granularity $N$ in \tool using CIFAR-10/ResNet-18. Results (Table~\ref{tab:impact_of_N}) show that the binary case ($N=2$, private vs. non-private) achieves strong accuracy ($75.98\%$ at $\epsilon=0.2$) with practical runtime (63,234s), comparable to baselines (FedAvg: 49,230s). Finer partitions ($N=3,4,5$) yield marginal accuracy gains ($\leq0.22\%$) but incur $>4\times$ runtime overhead due to exponential Shapley computation. Thus, $N=2$ is the recommended setting, balancing utility and efficiency. Meanwhile, \tool adds zero communication overhead over FedAvg; its cost is purely local computation.

\begin{table}[!t]
    \centering
    \caption{Impact of $N$ on \tool's performance when training ResNet-18 on CIFAR-10.}
    \scriptsize
    \setlength{\tabcolsep}{1.2pt}
    \begin{threeparttable}
    \begin{tabular}{c|c|c c| c c}
    \hline
     \multirow{3}{*}{Method} & \multirow{3}{*}{$N$} & \multicolumn{2}{c|}{$\epsilon=0.2$} & \multicolumn{2}{c}{$\epsilon=0.5$} \\ \cline{3-6}
     & &\makecell{Final-5\\ Average Acc.} & \makecell{Runtime (s)} & \makecell{Final-5 \\Average Acc.} & \makecell{Runtime (s)} \\\hline
     \multirow{4}{*}{\tool } & 2 &  75.98 $\pm$ 2.71  & 63,234 $\pm$ 1,415 & 82.14 $\pm$ 1.93 & 64,566 $\pm$ 1,689 \\

     & 3 &  75.87 $\pm$ 1.47  & 89,729 $\pm$ 3,619 & 82.04 $\pm$ 2.03 & 89,589 $\pm$ 2,479 \\

     & 4 &  75.83 $\pm$ 2.83  & 156,628 $\pm$ 2,987 & 81.86 $\pm$ 1.36 & 159,442 $\pm$ 3,147 \\

     & 5 &  75.76 $\pm$ 2.60  & 286,855 $\pm$ 3,861 & 81.67 $\pm$ 1.50 & 284,366 $\pm$ 3,383 \\\hline
    \end{tabular}
    \end{threeparttable}
    \label{tab:impact_of_N}
\end{table}

\section{Conclusion}
\label{sec:conclusion}


In this paper, we propose \tool, an explainable privacy mechanism for federated learning that calibrates Differential Privacy noise via Shapley Values, linking noise levels directly to the quantified contribution of private attributes. \tool provides rigorous privacy guarantees, converges under bounded noise, and significantly outperforms state-of-the-art methods in privacy-utility trade-off across standard benchmarks. Future work will focus on approximating Shapley Value computations to reduce complexity, enabling scalable, fine-grained privacy analysis with minimal accuracy loss for practical deployment.

\clearpage
\section{Acknowledgments}
This work was partly supported by the Natural Science Foundation of Shanghai (No. 23ZR1429600), the National Key R\&D Program of China (No. 2023YFB2704903), and Research and Development Program of the Department of Industry and Information Technology in Xinjiang Autonomous District (No. SA0304173).

\section{Compliance with Ethical Standards}
This is a numerical simulation study for which no ethical approval was required.

\bibliographystyle{IEEEbib}
\bibliography{refs}

\end{document}